\documentclass{article}

\usepackage[numbers,sort]{natbib}
\usepackage[final]{neurips_2020}

\usepackage[utf8]{inputenc} 
\usepackage[T1]{fontenc}    
\usepackage{hyperref}       
\usepackage{url}            
\usepackage{booktabs}       
\usepackage{amsfonts}       
\usepackage{nicefrac}       
\usepackage{microtype}      
\usepackage{graphicx}
\usepackage{cleveref}
\usepackage{subfig} 
\usepackage{multirow}
\usepackage{widetable}
\usepackage{hyperref}

\usepackage[noend]{algorithmic}
\usepackage[algoruled,vlined,nofillcomment,algo2e]{algorithm2e}
\usepackage{enumitem}
\newenvironment{protocol}[1][htb]
{
 \begin{algorithm2e}[#1]%
}{\end{algorithm2e}}
\newenvironment{protoenumerate}{%
\begin{enumerate}[itemindent=0em, labelwidth=1.2em, leftmargin=!, rightmargin=1.2em, itemsep=0ex]
}{%
\end{enumerate}
}

\usepackage{todonotes}

\makeatletter
\g@addto@macro{\UrlBreaks}{\UrlOrds}
\makeatother

\newcommand{\email}[1]{\href{mailto:#1}{\texttt{#1}}}

\title{Asymmetric Private Set Intersection with Applications to Contact Tracing and Private Vertical Federated Machine Learning}

\author{%
  Nick Angelou\\
  Morfix / OpenMined\\
  \email{angelou.nick@gmail.com}
  \And
  Ayoub Benaissa\\
  École Supérieure en Informatique, Sidi Bel Abbès / OpenMined\\
  \email{a.benaissa@esi-sba.dz}
  \And
  Bogdan Cebere\\
  Bitdefender / OpenMined\\
  \email{bogdan.cebere@gmail.com}
  \And
  William Clark\\
  OpenMined\\
  \email{will@willclark.tech}
  \And
  Adam James Hall\\
  Edinburgh Napier University / OpenMined\\
  \email{adam@openmined.org}
  \And
  Michael A. Hoeh\\
  apheris AI\\
  \email{m.hoeh@apheris.com}
  \And
  Daniel Liu\\
  University of California Los Angeles / OpenMined\\
  \email{daniel.liu02@gmail.com}
  \And
  Pavlos Papadopoulos\\
  Edinburgh Napier University / apheris AI\\
  \email{pavlos.papadopoulos@napier.ac.uk}
  \And
  Robin Roehm\\
  apheris AI\\
  \email{r.roehm@apheris.com}
  \And
  Robert Sandmann\\
  apheris AI\\
  \email{r.sandmann@apheris.com}
  \And
  Phillipp Schoppmann\\
  Humboldt-Universität zu Berlin / OpenMined\\
  \email{schoppmann@informatik.hu-berlin.de}
  \And
  Tom Titcombe\\
  Tessella / OpenMined\\
  \email{thomas.titcombe@tessella.com}
}

\begin{document}

\maketitle

\begin{abstract}
We present a multi-language, cross-platform, open-source library for asymmetric private set intersection (PSI) and PSI-Cardinality (PSI-C).
Our protocol combines traditional DDH-based PSI and PSI-C protocols with compression based on Bloom filters that helps reduce communication in the asymmetric setting.
Currently, our library supports C++, C, Go, WebAssembly, JavaScript, Python, and Rust, and runs on both traditional hardware (x86) and browser targets.
We further apply our library to two use cases: (i) a privacy-preserving contact tracing protocol that is compatible with existing approaches, but improves their privacy guarantees, and (ii) privacy-preserving machine learning on vertically partitioned data.
\end{abstract}

\section{Introduction}

In recent years, preserving privacy has became a fundamental requirement for any organization. Huge amounts of data are generated by social media, smart devices, or service providers (hospitals, banks, etc.), and analyzing them in a privacy preserving manner is not easy.  
Lately, privacy preserving techniques have emerged, including federated learning \cite{konevcny2016federated,yang2019federated}, secure multi-party computation \cite{goldreich1998secure}, or homomorphic encryption \cite{gentry2009fully,phong2018privacy}, which lead a much-needed privacy revolution.

Private Set Intersection (PSI) is a multi-party computation cryptographic protocol that allows two parties, each holding sets, to compute the intersection~\cite{freedman2004efficient,huang2012private,de2010practical,pinkas2018scalable,chase2020private} or the cardinality of the intersection~\cite{de2012fast,trieu2020epione} by comparing encrypted versions of these sets. 
Recently, several real-world applications of PSI and its variants have emerged.
\citet{DBLP:journals/iacr/IonKNPRSSSY19} applied a PSI-Sum protocol to advertising conversions aggregation, trying to solve the privacy problem between the ad supplier, which knows the users that have seen a particular ad, and the company, which knows who made a purchase and what they spent. \citet{DBLP:journals/iacr/BuddhavarapuKMS20} improved these results with a new library and implementation, Private-ID, with applications in randomized controlled trials or machine learning training.  
Other applications include private contact discovery~\cite{demmler2018contactdiscovery} or plagiarism detection~\cite{ihle2020PlagiarismDetection}.
Finally, there are several research implementations of PSI protocols~\cite{libPSI,chase2020private,pinkas2018scalable,kiss2017private,DBLP:conf/uss/KalesR0SW19}.
However, all of these have in common that they focus on a single setting, and only provide bindings for a single programming language.
In contrast, our goal is to enable the application of PSI in further use cases by providing an open-source library that is easy to use across language barriers and platforms.

\subsection{Contributions}

We present a versatile open-source library for asymmetric private set intersection (PSI) and PSI-Cardinality (PSI-C).
Our library combines the well-studied PSI protocol based on the decisional Diffie-Hellman (DDH) assumption with a Bloom filter compression to reduce the communication complexity.
Our library supports bindings for C++, C, Go, JavaScript, Python, and Rust, as well as multiple platforms including browser targets.
To the best of our knowledge, ours is the first PSI library to support such a wide set of languages.
Our implementation is available under an open source license at~\url{https://github.com/OpenMined/PSI}.

We describe the details of our architecture in \Cref{sec:architecture}.
Then, in \Cref{sec:applications}, the practicality of our library is exemplified with two use cases: privacy-preserving contact tracing (\Cref{sec:contact_tracing}) based on PSI-Cardinality, and secure matching of vertically partitioned data (\Cref{sec:vertical}) based on PSI.
Finally, in \Cref{sec:experiments}, we provide an experimental evaluation of our library.

\section{Protocol Description}

Protocol~\ref{proto} and Figure~\ref{fig:flow} in the appendix describe our variant of the DDH-based PSI and PSI-C protocols. The main difference to the presentation in~\citet[Figure 4]{trieu2020epione} is that our protocol uses a Bloom filter to compress the server's encrypted data set.
As observed in previous work~\cite{kiss2017private}, this can dramatically reduce communication in the asymmetric case.
Thus, our protocol can be seen as a hybrid of the works of \citet{trieu2020epione} and \citet{kiss2017private}.
We note that Bloom filters were chosen here for simplicity, and can be replaced by smaller data structures such as cuckoo filters~\cite{DBLP:conf/conext/FanAKM14} or Golomb-compressed sets~\cite{DBLP:journals/jea/PutzeSS09}. We also implement a more communication-efficient variant of our protocol using the latter. While we don't provide a full evaluation for that variant, we show the size differences between the data structures in Appendix~\ref{sec:gcs}.

\subsection{Security}
The security of our protocol follows directly from previous work. See for example \citet{de2012fast} for a security proof for the PSI-C variant without Bloom filters. Since our main difference is computing a Bloom filter with public parameters, which can be trivially simulated, security of Protocol~\ref{proto} follows.

Note that after the setup phase, communication only depends on $n$, the client's input size. Thus, our protocol is especially useful when the server set is static and can be reused across multiple queries.
For the PSI case, this can be done in a straight-forward way, since the server's secret key $k$ is never revealed.
For PSI-C, we need to be a bit more careful, since repeated queries may make it easier for the client to learn the elements in the intersection in addition to the intersection size.
For our contact tracing application (\Cref{sec:contact_tracing}), we therefore suggest to employ rate limiting, and rotate the server key by re-running the setup phase repeatedly.

\begin{protocol}[ht!]
    \caption{Our variant of the DDH-based PSI protocol. Adapted from~\citet[Figure 4]{trieu2020epione}.}\label{proto}
    \SetKwInput{KwInputs}{Inputs}
    \SetKwInput{KwParties}{Parties}
    \SetKwInput{KwParameters}{Parameters}
    
    \KwParameters{A cyclic group $G$ of order $p$, a hash function $H$ that maps inputs to $G$, and a boolean flag $\mathsf{RevealIntersection}$.}
    \KwInputs{Server: a set $X = \{x_1, \dots, x_N\}$, false-positive rate $p$.\newline Client: a set $Y = \{y_1, \dots, y_n\}$.}
    
    {\bfseries Server Setup:}\\
    \begin{protoenumerate}
        \item Server chooses a random key $k \gets \mathbb Z_p$.
        \item For each $i \in [N]$, the server computes $u_i = H(x_i)^k$.
        \item Server inserts $\{u_i \mid i\in [N]\}$ into a Bloom filter $\mathsf{BF}$ such that $n$ queries give a false positive with probability at most $p$, and sends $\mathsf{BF}$ to the client.
    \end{protoenumerate}
    
    {\bfseries Protocol:}\\
    \begin{protoenumerate}
        \item Client randomly samples $r \gets \mathbb Z_p$, and for each $y_i \in Y$ sends $m_i = H(y_i)^r$ to the server.
        \item For each $i\in [n]$, the server computes $m'_i = m_i^k$.
        \item If $\mathsf{RevealIntersection}$ is true, the server sends $\{m'_i \mid i \in [n]\}$ to the client ordered by $i$, otherwise ordered by $m'_i$.
        \item For each $i\in [n]$, the client computes $v_i = (m'_i)^{1/r}$.
        \item Client queries the Bloom filter for each $v_i$ and computes $S = \{i \in [n] \mid v_i \in \mathsf{BF}\}$. If $\mathsf{RevealIntersection}$ is true, the client outputs $S$, otherwise $|S|$.
    \end{protoenumerate}
\end{protocol}

\section{Architecture}
\label{sec:architecture}

The main focus of the library is on performance and versatility. To achieve that, the core of the library is built using C++, and our other language bindings all use this C++ core.
We use the elliptic curve P-256 to instantiate group $G$ in Protocol~\ref{proto}, relying on the implementation of \citet{DBLP:journals/iacr/IonKNPRSSSY19}.
For message serialization, we use Protocol Buffers~\cite{protobuf}, and our project is built using Bazel~\cite{bazel} to support builds across languages and platforms.

\Cref{fig:arch} in the appendix depicts our library's components and their interdependencies.
We use our C bindings to interface with languages like Go and Rust that don't have a native C++ interface.
For Python and JavaScript, we use the C++ core library directly.

\section{Applications}
\label{sec:applications}

\subsection{Privacy-Preserving Contact Tracing}
\label{sec:contact_tracing}

Smartphone based contact tracing allows to notify people who may have been exposed to infections, thereby allowing them to self-isolate or seek treatment. However contact tracing data poses risks of discrimination based on health status, or location leakage \cite{buchanan2020review}. Hence preserving the privacy of people's data is key for widespread adoption, especially considering that stopping a pandemic such as COVID-19 requires adoption of approximately 60\,\% of the whole population \cite{OxfordContactTracing}.

Several approaches for contact tracing have been developed this year, with similar patterns: Data of infected people is collected on a central server (e.g. by national health authority). Meanwhile people's smartphones exchange IDs between them and collect which other IDs they have been in contact with lately (e.g. based on Bluetooth) in a local list, regularly download a list of infected people's IDs from the server, and compute the intersection of both server and local ID lists to find out whether they have been exposed. Notable protocols include TCN  \cite{TCNCoalition} and DP3T \cite{DP3T}).

However, these protocols do not protect against linkage attacks, as essentially the server provides a list of infected IDs to the clients. This gives rise to the possibility of re-identifying infected persons, e.g. by recording IDs of people and using additional information (e.g., from cameras on the street). 

Linkage attacks can be mitigated by using PSI-C for computing the intersection. So instead of the server providing a list of infected IDs, only the intersection size is computed by following Protocol~\ref{proto}. Hence no list of infected IDs is provided by the server, rather both server and client encrypt their sets and jointly compute the intersection. 
Using PSI-C for contact tracing has also been proposed in concurrent work~\cite{trieu2020epione}.
To show the feasibility of our approach, and in particular its compatibility to existing protocols, we implement a privacy-preserving contact tracing library on top of the TCN protocol~\cite{TCNPSICardinality}.

\subsection{Privacy-Preserving Machine Learning on Vertically Partitioned Data}
\label{sec:vertical}

Vertical Federated Learning (VFL) \cite{yang2019federated,feng2020multi,liu2020asymmetrically} applies federated learning \cite{kairouz2019advances} to vertically distributed data, i.e., datasets that share partial information about the same entity, differing in the  features of each dataset. Different hospitals for example may have differing data about the same patient, but cannot simply merge this data across institutions due to privacy reasons. For such situations several approaches have been  many suggested in the past \cite{du2001privacy,du2004privacy,vaidya2002privacy,karr2009privacy,sanil2004privacy,wan2007privacy,gascon2017privacy}, including logistic regression \cite{hardy2017private,nock2018entity}.


One approach for such scenarios is Split Learning (SL) \cite{gupta2018distributed,vepakomma2018split,vepakomma2018peek} using a Split Neural Network (SplitNN), where the Neural Network (NN) is split among  participants, and each model segment acts as a self-contained NN. Each model segment trains and forwards its result to the next segment until completion. 
The security of SplitNN and its information leakage is being questioned \cite{kairouz2019advances}, but an enhanced privacy-preserving variant of SplitNN has been proposed in which the information leakage is reduced using distance correlation \cite{vepakomma2018supervised,vepakomma2019reducing,szekely2007}. Still, additional privacy-preserving methods could be incorporated in SL when dealing with very sensitive datasets.

In our work, we have realized a privacy-preserving implementation of SplitNN trained on vertically distributed data, called PyVertical \cite{PyVertical}. In our proof-of-concept (Appendix~\ref{pyverticalprocess}), we first utilise PSI for identifying matching entries in two vertically distributed datasets in a privacy preserving manner, and then train a SplitNN on this data, thereby  ensuring the privacy of the raw data. PyVertical is built upon the PySyft library \cite{ryffel2018generic}, which provides security features and mechanisms for training without sacrificing data privacy. Even though SplitNN does not provide formal security guarantees, PSI allows us to hide those data points that are not part of the intersection. Hence PSI is beneficial for just about any computation on vertically partitioned data (secure or not secure) if the parties want to hide elements that are not in the intersection.



\section{Evaluation and Conclusion}
\label{sec:experiments}

We benchmark our PSI library on an Amazon EC2 T3a.xlarge instance with 4 vCPUs at 2.5 GHz and 16 GiB memory.
Our results are presented in Table~\ref{table:experiments}.
It can be seen that there is little difference between languages that bind directly to our C++ core. This is to be expected, as the running time is dominated by elliptic curve operations, compared to which the overhead of crossing language barriers (including protobuf serialization) is small.
This changes when comparing WebAssembly and pure Javascript, which rely on cross-compilation of our library using Emscripten. Still, the overhead in modern browsers is less than 10x compared to the native version, and on the client side stays under 3 minutes for all sizes we tested.

The closest libraries to ours in previous work are the implementations of
\citet{kiss2017private} and \citet{DBLP:conf/uss/KalesR0SW19}.
Both focus on PSI for mobile applications, for example private contact discovery.
We compare our results with the ECC-NR-PSI protocol from \citet{DBLP:conf/uss/KalesR0SW19} for $N = 1M, n = 1k$. As noted there, this protocol already outperforms \citet{kiss2017private}.
As can be seen in Table~\ref{table:experiments}, the size of the setup message is smaller for \citet{DBLP:conf/uss/KalesR0SW19}. This is to be expected due to the fact that they use more efficient cuckoo filters~\cite{DBLP:conf/conext/FanAKM14}, and our library has an additional overhead due to our use of protocol buffers.
However, in both the online phase and in terms of total communication, our implementation improves on the results of \citet{DBLP:conf/uss/KalesR0SW19}.
In Appendix~\ref{sec:gcs} we also present experiments on a more communication-efficient variant of our protocol using Golomb-compressed sets.
We stress that the online times from~\cite{DBLP:conf/uss/KalesR0SW19} are not comparable to ours, since their experiments were run on different hardware, in particular using a smartphone for the client.

In conclusion, our results show that our library is highly competitive in terms of running time and communication.
At the same time, it is flexible enough to support multiple platforms and languages, including browsers.
Possible improvements can be made by reducing the setup communication, for example using cuckoo filters~\cite{DBLP:conf/conext/FanAKM14,DBLP:conf/uss/KalesR0SW19}.
Finally, we see extensions like PrivateID~\cite{DBLP:journals/iacr/BuddhavarapuKMS20} as promising future work.

\begin{table}[t]\scriptsize
\caption{\label{tab:benchmarks}Benchmarks in seconds for $n$ client elements and $N = 1M$ server elements. The probability of a false positive over $n$ lookups was set to $p = 10^{-9}$.
An asterisk ($^*$) indicates experiments that were run on a smaller server set ($N = 100k$) and extrapolated from there. 
Numbers for previous work were taken from \cite[Table 6]{DBLP:conf/uss/KalesR0SW19}.
Cells with dagger ($^\dag$) were summed up in~\cite{DBLP:conf/uss/KalesR0SW19}, so we only report the sum.
}\label{table:experiments}
\begin{widetable}{\textwidth}{llcccccccccc}
\toprule


 Operation & Size & C++ & C & Go & Python & WebAssembly & JS & \cite{DBLP:conf/uss/KalesR0SW19} & Comm. & \cite{DBLP:conf/uss/KalesR0SW19} (Comm.) \\

\midrule
\multirow{3}{*}{Server: Setup}  
& $n = 1k$ & 184.1 & 181.5 & 183.9 & 188.9 & 1573.8$^*$ & 9128$^*$ & 241.54 & 6.85 MiB & 4.19 MiB \\
& $n = 10k$ & 184.2 & 181.7 & 184.0 & 188.4 & 1571.8$^*$ & 9273$^*$ & - & 7.42 MiB & - \\
& $n = 100k$ & 184.2 & 181.8 & 184.3 & 189.1 & 1573$^*$ & 9139.1$^*$ & - & 7.99 MiB & - \\
\midrule
\multirow{3}{*}{Client: Request}   
& $n = 1k$ & 0.18 & 0.17 & 0.18 & 0.185 & 1.57 & 9.1 & 2.92$^\dag$ & 34.18 KiB & 2.00 MiB\\
& $n = 10k$ & 1.8 & 1.77 & 1.8 & 1.870 & 15.6 & 92 & - & 341.79 KiB & -\\
& $n = 100k$ & 18.09 & 17.8 & 18.2 & 18.2 & 156.8 & 909.4 & - & 3.33 MiB & - \\
\midrule
\multirow{3}{*}{Server: Response}  
& $n = 1k$ & 0.11 & 0.11 & 0.17 & 0.115 & 1.2 & 6.96 & $^\dag$ & 34.17 KiB & 4.07 MiB \\
& $n = 10k$ & 1.13 & 1.13 & 3.02 & 1.154 & 12 & 70.8 & - & 341.79 KiB & - \\
& $n = 100k$ & 11.3 & 11.3 & 29.5 & 11.5 & 120.9 & 701 & - & 3.33 MiB & -\\
\midrule
\multirow{3}{*}{Client: Intersection} 
& $n = 1k$ & 0.11 & 0.11 & 0.9 & 0.120 & 1.2 & 6.97 & $^\dag$ & - & - \\
& $n = 10k$ & 1.17 & 1.16 & 4.6 & 1.193 & 12.1 & 71.39 & - & - & -\\
& $n = 100k$ & 11.8 & 11.6 & 41.6 & 11.9 & 122 & 710 & - & - & -\\
\bottomrule
\end{widetable}
\end{table}

\section*{Acknowledgements}
Realizing both the PSI library and the two implementations for contact tracing and vertically partitioned federated learning has only been possible with the help of many people who contributed to this. We extend our deepest gratitude to everybody that has been and continues to be involved in this effort.

\bibliographystyle{plainnat}
\bibliography{references}

\appendix

\section{Protocol Flow and Library Architecture}

\begin{figure} 
  \centering
  \includegraphics[width=0.7\textwidth]{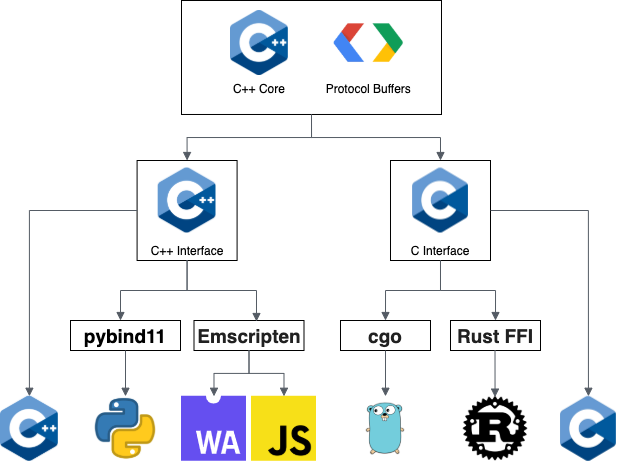}
  \caption{PSI library high-level architecture and language dependencies}
  \label{fig:arch}
\end{figure}

\begin{figure} 
  \centering
  \includegraphics[width=0.8\textwidth]{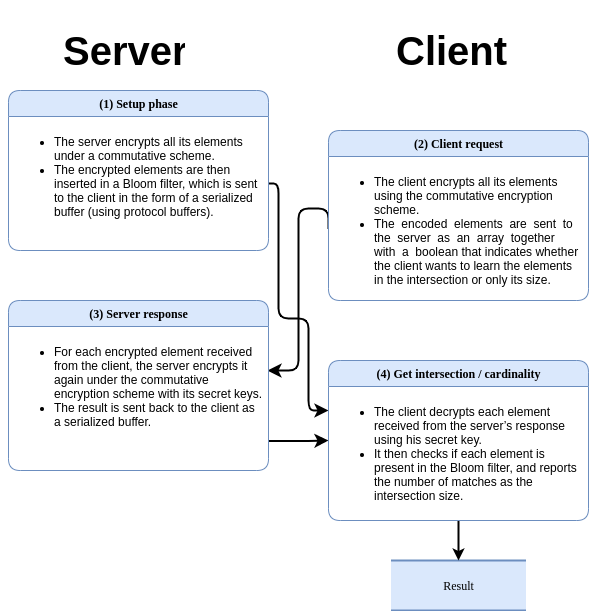}
  \caption{PSI library high-level flow.}
  \label{fig:flow}
\end{figure}

PSI library high-level flow can be seen in Figure~\ref{fig:flow}. PSI high-level architecture and the associated languages dependencies can be seen in Figure~\ref{fig:arch}.

\section{Communication Size}
\label{sec:gcs}
For many use cases of PSI, the server must send a large set of setup data to the client. Therefore, in addition to Bloom filters, we also implement Golomb coding~\cite{DBLP:journals/jea/PutzeSS09} for better compression of the setup set. We will refer to this data structure as Golomb-compressed sets (GCS). We compare the sizes of Golomb-compressed sets and Bloom filters that needs to be sent to the client in Table~\ref{table:comm}. Internally, the GCS implementation uses a one-pass encoding scheme on the server side, and an on-the-fly decoding scheme for bulk intersection on the client side, so it is similar in speed to Bloom filters for bulk insertion and intersection. However, Bloom filters are faster at random accesses and insertions of a small number elements. Therefore, it is important to weigh the costs of communication with the costs of random access lookups when deciding which underlying data structure to use.

\begin{table}[t]\small
\centering
\caption{\label{table:comm}Communication sizes for Bloom filters, Golomb-compressed sets, and naively storing 64-bit integers with varying false-positive rates. We insert $10^4$ elements and calculate the size of each data structure.}
\begin{tabular}{lccc}
\toprule

FPR & Naive & Bloom filter & GCS\\

\midrule
$10^{-6}$ & 80 KB & 36 KB & 27 KB\\
$10^{-7}$ & 80 KB & 42 KB & 31 KB\\
$10^{-8}$ & 80 KB & 48 KB & 35 KB\\
$10^{-9}$ & 80 KB & 54 KB & 39 KB\\
$10^{-10}$ & 80 KB & 60 KB & 43 KB\\
$10^{-11}$ & 80 KB & 66 KB & 48 KB\\
$10^{-12}$ & 80 KB & 72 KB & 52 KB\\
\bottomrule
\end{tabular}
\end{table}

\section{PyVertical proof-of-concept process}
\label{pyverticalprocess}

\begin{figure}[h!]
    \centering
    \subfloat[Full dataset]{{\includegraphics[width=5.6cm]{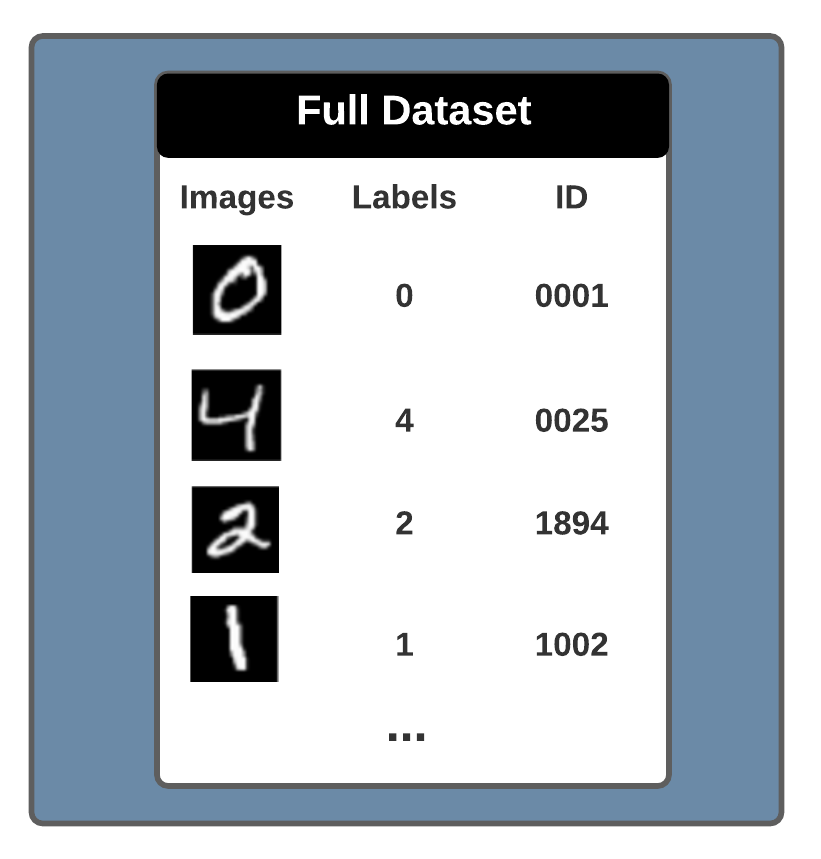} }}%
    \qquad
    \subfloat[Split images and labels datasets]{{\includegraphics[width=7.1cm]{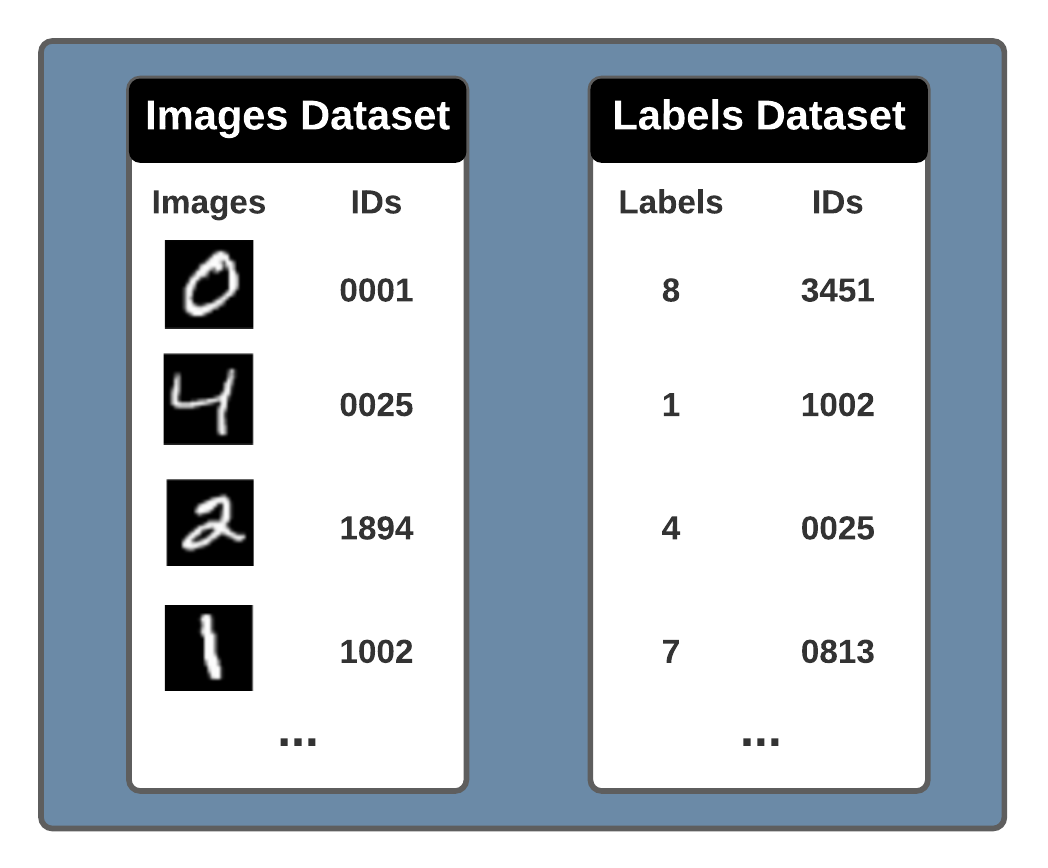} }}
    \qquad 
    \subfloat[PSI linkage and ordering]{{\includegraphics[width=6cm]{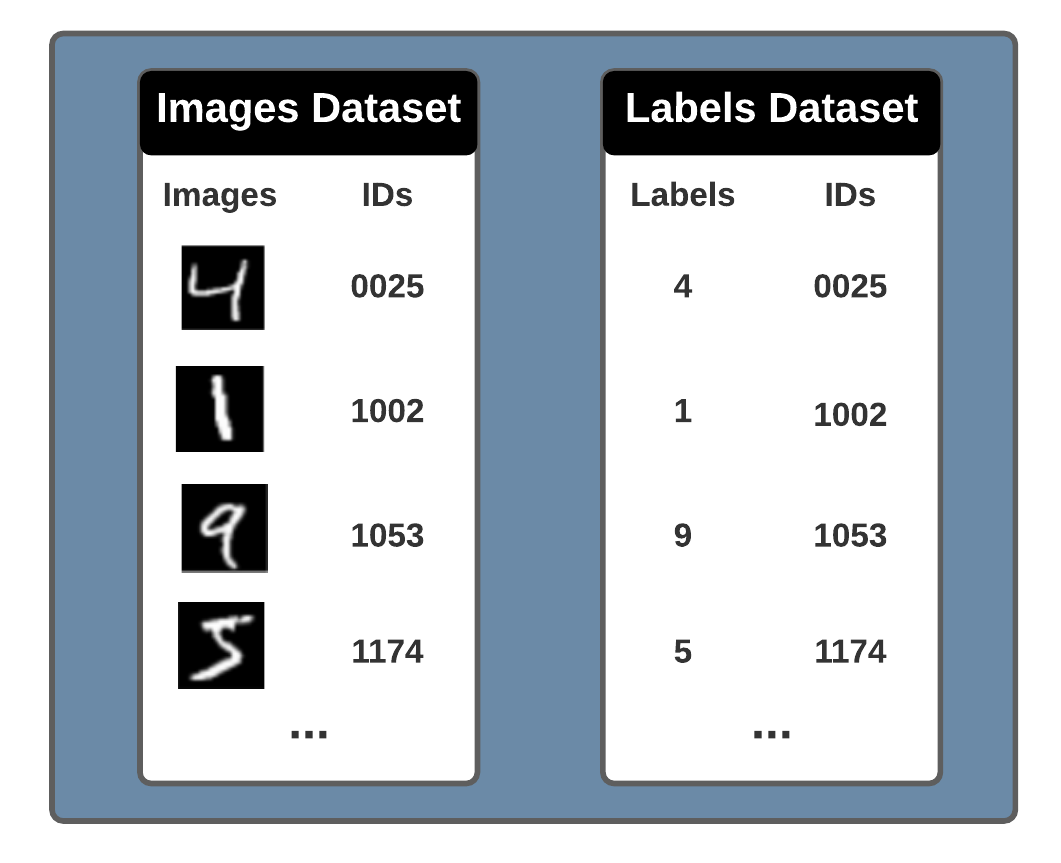} }}%
    \qquad
    \subfloat[SplitNN training]{{\includegraphics[width=7cm,height=4.85cm]{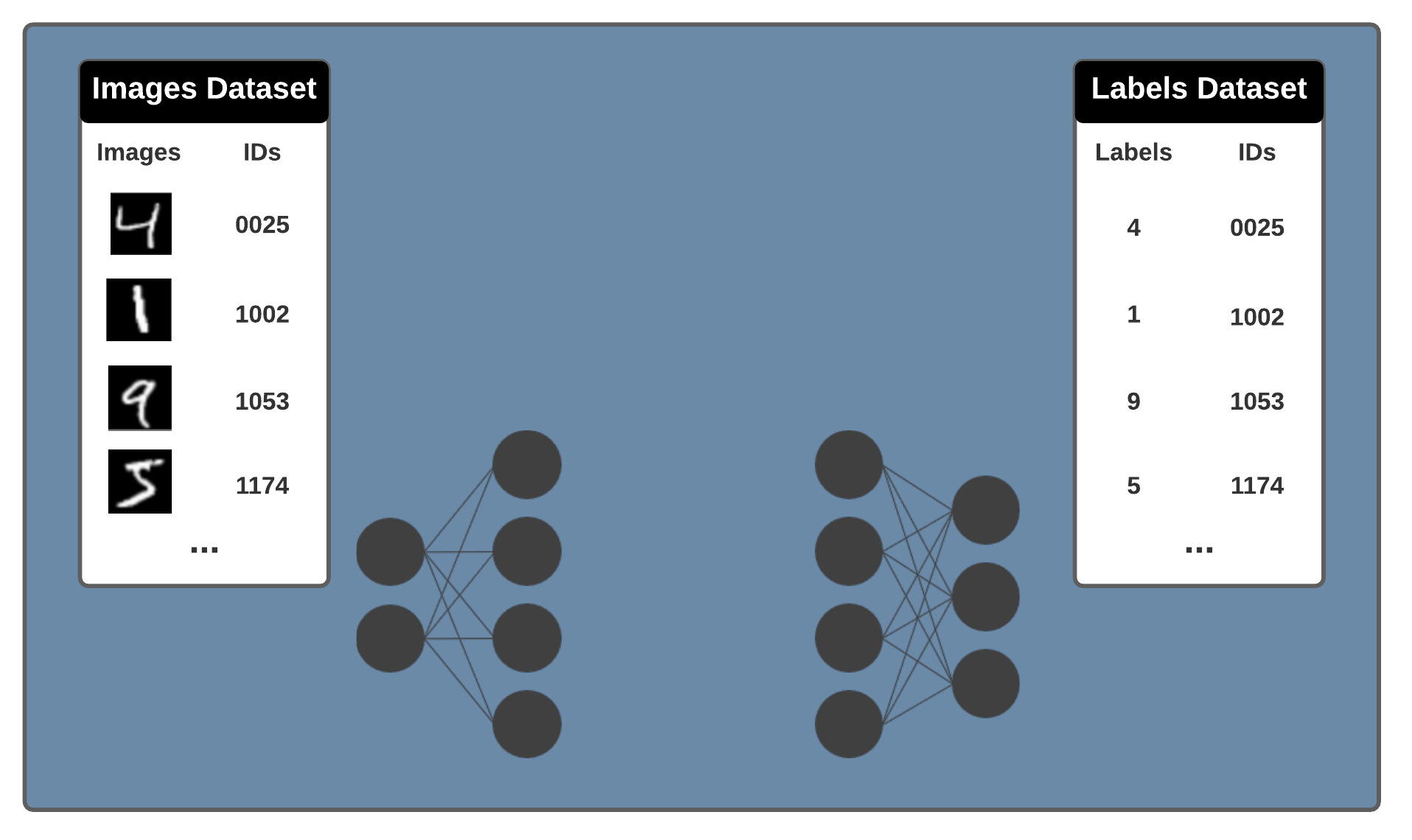} }}
    \caption{PyVertical proof-of-concept implementation}
    \label{fig:pyvertical}
\end{figure}

For our proof-of-concept of PyVertical we used the MNIST dataset of handwritten images and their labels \cite{deng2012mnist}. The overall process is shown in  Figure~\ref{fig:pyvertical}. MNIST is a horizontally distributed dataset. 
As shown in Figure~\ref{fig:pyvertical}, a) we have added a new \textit{IDs} field to this data which assigns unique IDs to each data point. Furthermore, as shown in Figure~\ref{fig:pyvertical}, b), we split the full dataset into two datasets; one with the handwritten images and their IDs, and one with the labels and their IDs. In the next step, we shuffle the datasets and randomly remove data points from both. 

In the actual experiment, we utilise PSI to identify matching IDs in the two datasets, as shown in Figure~\ref{fig:pyvertical}, c), hence link the data points with shared IDs, and order both datasets accordingly. Data points that are not elements of the intersection are being purged in this example. Finally, we create a SplitNN, as shown in Figure~\ref{fig:pyvertical}, d),  where one part of the network is trained on the images dataset and the other part of the network is trained on the labels dataset. This allows to train the model in a privacy-preserving manner by leveraging PySyft's pointers functionality \cite{ryffel2018generic} which keeps the raw training data hidden throughout the training process. 


\end{document}